\documentstyle[12pt,amssymb,amsfonts,amsbsy]{article}
\begin{document}
\author{Ilja Schmelzer\thanks
       {WIAS Berlin}}

\title{Post-Relativistic Gravity --- A Hidden Variable Theory For
General Relativity}

\begin{abstract}
\sloppypar
Post-relativistic gravity is a hidden variable theory for general
relativity. It introduces the pre-relativistic notions absolute space,
absolute time, and ether as hidden variables into general relativity.
Evolution is defined by the equations of general relativity and the
harmonic coordinate condition interpreted as a physical equation.
There are minor differences in predictions compared with general
relativity (i.e. trivial topology of the universe is predicted).

The unobservable absolute time is designed to solve the problem of
time in quantization of general relativity. Background space and time
define a Newtonian frame for the quantization of the gravitational
field. By the way, a lot of other conceptual problems of quantization
will be solved (i.e. no constraints, no topological foam, no black
hole and bib bang singularities, natural vacuum definition for quantum
fields on classical background).

\end{abstract}

\maketitle

\section{Introduction}

Post-Relativistic Gravity (PG) is a hidden variable theory of General
Relativity (GR). The hidden variables are an absolute, {\it true time}
t and an absolute {\it background space} with affine structure. This
space is filled with an {\it ether} described by the field
$g_{ij}(x,t)$.  The interaction of the ether with material fields is
described by the the Einstein equations

\[R_{ik} - {1\over 2} g_{ik}R = {8\pi k \over c^4} T_{ik} \]

and the connection with the hidden variables by the harmonic
coordinate equations

\[ \Gamma^k = g^{ij} \Gamma^k_{ij} = 0. \]

The name {\bf post-relativistic} for the theory seems natural to
describe the revival of Newtonian notions in the relativistic context.

PG doesn't make further assumptions about the internal structure of
the ether, nonetheless there are some natural guesses: The ether
velocity can be defined by $v^i = g^{i0}/g^{00}$.  The space part
$g_{ij}(x), i,j=1,2,3;$ reminds a deformation tensor. The remaining
scalar step of freedom requests an interpretation in terms of speed of
light. The ether is not stationary, thus one conceptual problem of
classical ether theory is solved in PG by unification with gravity.

Considering the interaction of the ether with clocks, which is
described by the Einstein equations, we can conclude, as in GR, that
the result of clock measurement is proportional to the notion of
GR proper time 

\[\tau = \int \sqrt{g_{ij}dx^i dx^j} \]

True time t remains not measurable, but is considered to be the more
fundamental notion of time which describes past, present, future and
causality.  GR clock time $\tau$ describes only the distortion of
clock measurement caused by gravity.  The absense of an appropriate
time measurement is considered as a natural property of time we can
find in non-relativistic quantum mechanics too \cite{Isham}.

Affine space and time in PG lead to a finite-dimensional symmetry
group.  Using the Noether theorem we obtain local energy and momentum
densities known in GR as the energy-momentum pseudotensor.  It's
coordinate-dependence obtains a natural interpretation as the
dependence of the hidden steps of freedom, especially the ether
velocity.

PG in itself is interesting enough already from pure methodological
reasons. It shows that a completely different metaphysical
interpretation of our universe including absolute time and an ether is
compatible with GR. But the much more interesting point is the
essential simplification of quantization.

\section{Historical Context}

PG was developed by I. Schmelzer \cite{Schmelzer}.

PG can be considered as a generalization of the Lorentz-Poincare
version of special relativity \cite{Poincare} to general
relativity. It is remarkable that the distinction between two notions
of time --- true time and time measurement with clocks --- has been
introduced already by Newton \cite{Newton}.

The harmonic coordinate equation has been often used in GR, starting
with Lanczos \cite{Lanczos} and Fock \cite{Fock}, up to actual
attempts to use them as gauge conditions or clock fields in
quantization.  But in the context of GR they cause some problems like
different solutions for the same metric and solutions which don't
cover the whole solution.

Logunov et.al. have introduced the harmonic coordinate equation as a
physical equation into their {\it Relativistic Theory of Gravity}
\cite{Logunov}, \cite{Vlasov} and found the related modification for
the black hole and big bang scenario. Different from PG they have
introduced a Minkowski background, and their argumentation was based
on incorrect criticism of GR \cite{Seldovich}.

The Isham-Kuchar approach \cite{Kuchar} interpretes harmonic space and
time coordinates as gravity-coupled massless fields used to identify
instants of time and points in space. This approach remains fully
inside standard GR, a quantization based on this approach faces the
same problems as usual GR quantization. Especially, a ``clock field''
t will be uncertain and measurable, different from quantum mechanical
time and PG true time.

PG may be described in the classification give by Isham \cite{Isham}
as ``GR forced into a Newtonian framework''.  Isham mentions the
reduction of the symmetry group in such an approach we find in PG
too. But the reason for rejection given there --- ``theoretical
physicists tend to want to consider all possible universes under the
umbrella of a single theoretical structure'' --- is not valid for PG,
because PG describes in their context all possible universes.
\footnote{Isham mentiones \cite{Valentini} as a ``recent suggestion
that a preferred foliation of spacetime could arise from the existence
of nonlocal hidden-variables.'' --- an idea which seems close to PG.}

\section{Comparison With General Relativity}

Despite the completely different metaphysics of PG it is de-facto not
possible to distinguish GR and PG by a classical experiment.
Nonetheless PG is a different theory which makes additional
predictions compared with GR.

First, topologically nontrivial solutions are obviously forbidden in
PG.  Another difference is completeness. A complete PG solutions has
to be defined for all x and t, which is different from completeness of
the metric $g_{ij}(x,t)$. Thus, the underlying GR solution of a
complete PG solution may be incomplete from GR point of view.  A
trajectory with finite ``proper time'' for infinite true time doesn't
lead to conceptual problems because it is simply interpreted as
``freezing'' of physical effects caused by an extremal field.

The most interesting example is the black hole collaps. The complete
PG solution here is the part of the GR solution before horizon
formation. 
\footnote{This adds evidence that the semiclassical
approximation breaks down near the black hole horizon \cite{Keski}.}
The PG solution for the big bang is also complete for $t \to -\infty$.
On the other hand, the reverse may be true too. A complete GR solution
may be incomplete from PG point of view.

The other essential difference is symmetry. The symmetry group of PG
is an affine variant of the Galilei group --- a finite dimensional
group which includes translations in space and time. A minor
consequence is a preference of the flat universe solution in PG. The
curved universes are still possible, but no longer homogeneous.
Another consequence is the possibility of definition of a local
energy-momentum tensor using the Noether theorem. We obtain the
well-known energy-momentum pseudotensor:

\begin{eqnarray*}
t^{ik} &= &{c^4 \over 16\pi k} \left\{
(2\Gamma^n_{lm}\Gamma^p_{np} -
  \Gamma^n_{lp}\Gamma^p_{mn} -
  \Gamma^n_{ln}\Gamma^p_{mp}) (g^{il}g^{km} - g^{ik}g^{lm})\right. + \\
&+ &g^{il}g^{mn}(
  \Gamma^k_{lp}\Gamma^p_{mn} +
  \Gamma^k_{mn}\Gamma^p_{lp} -
  \Gamma^k_{np}\Gamma^p_{lm} -
  \Gamma^k_{lm}\Gamma^p_{np})  + \\
&+ &g^{kl}g^{mn}(
  \Gamma^i_{lp}\Gamma^p_{mn} +
  \Gamma^i_{mn}\Gamma^p_{lp} -
  \Gamma^i_{np}\Gamma^p_{lm} -
  \Gamma^i_{lm}\Gamma^p_{np}) + \\
&+ &\left. g^{kl}g^{mn}(
  \Gamma^i_{ln}\Gamma^k_{mp} -
  \Gamma^i_{lm}\Gamma^k_{np}) \right\}\\
\end{eqnarray*}

The coordinate-dependence of the pseudo-tensor in GR has a natural
explanation in PG as the dependence of energy and momentum from the
hidden velocity of the ether.  Full local energy and momentum
densities are simply ``hidden variables'' in GR too --- a natural
point of view, because they are dual to the coordinates.

The additional predictions of PG are in good agreement with
observation. We have not yet seen any topologically nontrivial
solution, and the universe seems to be approximately
flat. Nonetheless, this cannot be considered as an experimental
verification of PG. At the classical level PG and GR are de-facto
identical in their experimental predictions.

\section{Quantization of Post-Relativistic Gravity}

PG and GR may be distinguished during quantization. It is essentially
to remark that quantum PG is a quantum theory with GR as the classical
limit, but is not quantum GR and that's why avoids most of the
conceptual problems of GR quantization. It is using the hidden
Newtonian structure as a frame for classical quantization.

This solves the ``problem of time'' in quantum gravity. The concepts
of time of GR and non-relativistic quantum mechanics are in
contradiction (\cite{Isham}, \cite{Kitada}, \cite{Au}). ``Whatever the
final version of quantum gravity is like, as long as the requirement
of general covariance is imposed, the quantum theory would seem to
have the problem of time'' \cite{Alwis}. In quantum PG these two
concepts of time are simply defined by two different objects --- true
time t and clock time $\tau$. By the way, this solves
causality-related problems like ``uncertain causality'' caused by
uncertain gravitational field, superluminal tunneling
\cite{Steinberg}, \cite{Chiao}, Bell's inequality \cite{Bell}.

Together with time and causality, the background space in quantum PG
is external and fixed. Obviously this allows to define a position
operator, which is meaningless in GR where we have even different
topologies. But other observables need this space too. Indeed, after
an interaction of a state $|g^1\rangle+|g^2\rangle$ with a test
particle $|\psi\rangle$ we obtain a state
$|\psi^1\rangle\otimes|g^1\rangle+|\psi^2\rangle\otimes|g^2\rangle$,
and the probability that this interaction has changed the initial
superpositional state into $|g^1\rangle-|g^2\rangle$ depends on the
scalar product $\langle\psi^1|\psi^2\rangle$. These two states
$|\psi^1\rangle$ and $|\psi^2\rangle$ are the results of evolution on
different gravitational background, thus in the semiclassical
approximation functions on different solutions. In GR, such a scalar
product of functions defined on different solutions is undefined. PG
defines it straightforward using the background space.

The situation is similar for other observables too: Local energy and
momentum density are not measurable already in classical GR, not to
talk about quantum theory. In PG this problem doesn't exist.

The number of particles of a quantum field is already problematic for
a quantum field on a fixed GR background --- the definition of the
vacuum state is not diffeomorphism-invariant.  In PG it is possible to
define a scheme which allows to define a unique, ``natural'' vacuum
state depending on the field configuration at a given moment of true
time t.

Thus, all usual classical observables may be easily introduced into
quantum PG, but lead to conceptual problems in quantum GR.  But there
are also other advantages of quantum PG:

 \begin{itemize}

 \item PG is a theory which is ``relativistically but not Lorentz
invariant'' in Bell's classification \cite{Bell}. Thus, the violation
of Bell's inequality may be described by real, hidden mechanisms which
violate Einstein's causality, but not PG causality. Quantum state
reduction occurs in the ``preferred frame'' defined by t = const (See
also Cohen and Hiley \cite{Cohen}).

 \item The tetrade mechanism may be modified using the subdivision
into space and time which is already fixed in PG. This reduction into
a triade formalism reduces the resulting gauge group from SO(3,1) to
SO(3).  The compactness of SO(3) is a great technical advantage
(i.e. for lattice gauge theory \cite{Baez}).

 \item Because we have well-defined translational symmetry and local
energy and momentum densities, the Hamilton formulation is much
easier, the canonical quantization scheme doesn't lead to constraints
but to a normal evolution equation.

 \item We obviously have no problems with topology --- no topological
foam, no wormholes.

 \item The different picture of the black hole collaps without an
actual part behind the horizon solves the problems related with
conservation laws \cite{Wheeler}.

 \end{itemize}

Full quantization of a nonlinear field theory like PG remains to be a
complex problem. But PG allows to solve in a straightforward way many
of the fundamental, conceptual problems of GR quantization. It seems
that the remaining problems of PG quantization have technical, not
conceptual character.

\end{document}